2015-08-14

# The Diagonal Lemma Fails in Aristotelian Logic

*X.Y. Newberry*

Gödel's theorem does not need an introduction. But it is generally not acknowledged that all instances of Gödel's sentence are vacuous (see below). On the contrary Aristotelian logic permits only sentences that are *not* vacuous. The consequences of these observations are surprising.

What does the article title mean? First of all by "Aristotelian logic" we will understand the logic of the traditional syllogism. By "classical logic" we will mean modern symbolic logic. It is well known that the laws of Aristotelian logic are valid only if all the terms refer to non-empty sets.

Traditional Aristotelian logic recognizes four types of sentences

| | |
|---|---|
| A: | All F are G |
| E: | No F is G |
| I: | Some F are G |
| O: | Some F are not G |

Certain relationships are said to hold between these types. For example A and E are *contraries*. It means that both cannot be true, but both can be false. Here the modern logicians spotted a problem. Suppose that we interpret the four types as in Table 1 below.

| | |
|---|---|
| A: | $\sim(\exists x)(Fx \& \sim Gx)$ |
| E: | $\sim(\exists x)(Fx \& Gx)$ |
| I: | $(\exists x)(Fx \& Gx)$ |
| O: | $(\exists x)(Fx \& \sim Gx)$ |

Table 1

Suppose further that $F$ is empty, i.e. that $\sim(\exists x)Fx$. Then according to classical logic both A and E will be true. The above interpretation [Table 1] does not hold. The issue was analyzed by P. F. Strawson. (Strawson, 1952, pp. 163-179) He showed that given the interpretation below

A: $\sim(\exists x)(Fx \;\&\; \sim Gx) \;\&\; \boldsymbol{(\exists x)Fx \;\&\; (\exists x)\sim Gx}$

E: $\sim(\exists x)(Fx \;\&\; Gx) \;\&\; \boldsymbol{(\exists x)Fx \;\&\; (\exists x)Gx}$

I: $(\exists x)(Fx \;\&\; Gx) \lor \boldsymbol{\sim(\exists x)Fx \lor \sim(\exists x)Gx}$

O: $(\exists x)(Fx \;\&\; Gx) \lor \boldsymbol{\sim(\exists x)Fx \lor \sim(\exists x\sim)Gx}$

Table 2

all the laws of the traditional syllogism will hold. (Strawson, 1952, p. 173) Traditional logic assumes that the subject term refers to something that does exist. However, the formulae in Table 2 are implausible translations of the natural language sentences. (Strawson, 1952, p. 173) So he proposed to take the term $(\exists x)Fx$ as a *presupposition*. It means that $\sim(Ex)Fx$ does not imply that A is false, but rather $(Ex)Fx$ "is a necessary precondition not merely of of the truth of what is said, but of its being *either* true *or* false." [Original italics] (Strawson, p. 174) We will, however, do one better and take the entire $(\exists x)Fx \;\&\; (Ex)\sim Gx$ as the presupposition. Then A *is neither true nor false if* $(\exists x)Fx \;\&\; (\exists x)\sim Gx$ is not true. For our purposes it is important to note that there is no such thing as a vacuously true proposition. Vacuous propositions are by definition neither true nor false.

Such a logic *can* be formalized. This can be accomplished by generalizing truth-relevant logic (Diaz, 1981) to the predicate calculus. In this logic the sentences

$(P \;\&\; \sim P) \rightarrow Q$

$\sim(P \;\&\; \sim P) \lor Q$

$\sim((P \;\&\; \sim P) \;\&\; \sim Q)$

are not *truth-relevant tautologies* (Diaz, 1981, p. 67.) Similarly in Strawson's logic the sentences

$(x)(Fx \rightarrow Gx)$
$(x)(\sim Fx \vee Gx)$
$\sim(\exists x)(Fx \,\&\, \sim Gx)$

are not true if $\sim(\exists x)Fx$. (Nor are they false.)

The author of truth-relevant logic probably never realized that his system was a propositional counterpart of the traditional Aristotelian logic! He arrived at it from a different angle, the angle of relevance. But truth-relevant logic *can* be extended not only to monadic predicate calculus but also to the logic of relations. (Newberry, 2014)

* * * * *

Let us now turn our attention to the diagonal lemma and in particular to Gödel's theorem. In Peano Arithmetic there exists a decidable relation $Diag(y,z)$ such that if $y$ is the Gödel number of a formula with one free variable then $z$ is the Gödel number number of the formula obtained from $y$ by substituting (the numeral of) the Gödel number of $y$ for the free variable in $y$. Further let $Prf(x,z)$ be a predicate such that $x$ is the Gödel number of a sequence that is a proof of the sentence with Gödel number $z$. Then consider the formula

$$\sim(\exists x)(\exists z)(Prf(x,z) \,\&\, Diag(y,z)) \tag{U}$$

with one free variable $y$. Let the constant $k$ be the Gödel number of U. We substitute $k$ for the free variable $y$ in U. We obtain

$$\sim(\exists x)(\exists z)(Prf(x,z) \,\&\, Diag(k,z)) \tag{G}$$

As a result of this construction $Diag(k,z)$ is satisfied only by the Gödel number of G. We will denote the Gödel number of G as '$<G>$'. Then according to classical logic G is equivalent to

$$\sim(\exists x)Prf(x, <G>) \qquad \text{(H)}$$

and thus

$$\sim(\exists x)(\exists z)(Prf(x,z)\ \&\ Diag(k,z)) \leftrightarrow\ \sim(\exists x)Prf(x,<G>) \qquad \text{(J)}$$

The sentence (J) above is an instance of the diagonal lemma also known as the fixed point theorem. We replaced the free variable $z$ in $\sim(\exists x)Prf(x,z)$ with the Gödel number of some sentence $\varphi$ such that $\varphi \leftrightarrow\ \sim(\exists x)Prf(x,<\varphi>)$. In this case $\varphi$ happens to be G.

Now we are coming to the crux of the matter. Let us "unroll" G:

$$\sim(\exists x)(Prf(x,1)\ \&\ Diag(k, 1))$$
$$\sim(\exists x)(Prf(x,2)\ \&\ Diag(k, 2))$$
$$\sim(\exists x)(Prf(x,3)\ \&\ Diag(k, 3))$$
$$\ldots$$

For any $n$ either $\sim(\exists x)(Prf(x,n)$ or $\sim(\exists x)Diag(k, n)$. Given the following equalities

$$\sim(\exists x)(Prf(x,n)\ \&\ Diag(k,n)) \Leftrightarrow$$
$$(x)(Prf(x,n) \rightarrow\ \sim Diag(k,n)) \Leftrightarrow$$
$$(x)(Diag(k,n) \rightarrow\ \sim Prf(x,n))$$

we find that if $n = <G>$ then

$$(x)(Prf(x,n) \rightarrow\ \sim Diag(k,n))$$

is vacuous, else if $\sim(n = <G>)$ then

$$(x)(Diag(k,n) \rightarrow\ \sim Prf(x,n))$$

is vacuous. So let $n = <G>$:

$$\sim(\exists x)(\boldsymbol{Prf(x,<G>)} \;\&\; Diag(k,<G>)) \qquad\qquad (K)$$

According to the logic of presuppositions both terms in K must refer to non-empty sets. In particular $(\exists x)Prf(x,<G>)$ must hold; it is a *presupposition* of K. That is, *if* $\sim(\exists x)Prf(x,<G>)$ *then K cannot be true!* The equivalence J no longer holds. By cutting the Gordian knot we are able to say that G is not true even though G cannot say it of itself.

# Bibliography

Diaz, M.R. (1981) *Topics in the Logic of Relevance*, Munich, Germany: Philosophia Verlag.

Newberry, X.Y. (2014) Generalization of the Truth-relevant Semantics to the Predicate Calculus, http://arxiv.org/ftp/arxiv/papers/1509/1509.06837.pdf

Strawson, P.F. (1952) *Introduction to Logical Theory*, London, Methuen